\documentclass[preprint]{aastex}
\usepackage{lscape} 
\usepackage{amsmath, amsthm,amssymb}
\usepackage{multirow}
\usepackage{url}

\title{The Hubble Wide Field Camera 3 Test of Surfaces in the Outer Solar System: Spectral Variation on Kuiper Belt Objects}

\author{Wesley C. Fraser {$^{1,2}$}}
\altaffiltext{1}{Herzberg Institute of Astrophysics, 5071 W. Saanich Rd. Victoria, BC V9E 2E7}
\email{wesley.fraser@nrc.ca}
\author{Michael E. Brown {$^2$}}
\altaffiltext{2}{California Institute of Technology, 1200 E. California Blvd. Pasadena, CA 91101}
\author{Florian Glass {$^{3,4}$}}
\altaffiltext{3}{Observatoire de Genve, UniversitŽ de Genve, 51 chemin des Maillettes, CH-1290 Sauverny, Switzerland}
\altaffiltext{4}{Dept. of Physics and Astronomy, University of Victoria, Victoria, BC V8W 3P6, Canada}

\date{} 

\slugcomment{Submitted to ...}
\received{É}
\revised{É}

\begin{abstract}
Here we present additional photometry of targets observed  as part of the Hubble Wide Field Camera 3 Test of Surfaces in the Outer Solar System. 12 targets were re-observed with the Wide Field Camera 3 in optical and NIR wavebands designed to compliment those used during the first visit. Additionally, all observations originally presented by Fraser and Brown (2012) were reanalyzed through the same updated  photometry pipeline. A reanalysis of the optical and NIR colour distribution reveals a bifurcated optical colour distribution and only two identifiable spectral classes, each of which occupies a broad range of colours and have correlated optical and NIR colours, in agreement with our previous findings. We report the detection of significant spectral variations on 5 targets which cannot be attributed to photometry errors, cosmic rays, point spread function or sensitivity variations, or other image artifacts capable of explaining the magnitude of the variation. The spectrally variable objects are found to have a broad range of dynamical classes, absolute magnitudes, exhibit a broad range of apparent magnitude variations, and are found in both compositional classes. The spectrally variable objects with sufficiently accurate colours for spectral classification maintain their membership, belonging to the same class at both epochs. 2005 TV189 exhibits a sufficiently broad difference in colour  at the two epochs that span the full range of colours of the neutral class. This strongly argues that the neutral class is one single class with a broad range of colours, rather than the combination of multiple overlapping classes.
\end{abstract}

\begin{document}

\maketitle

\section{Introduction \label{sec:Intro}}
Small Kuiper Belt Objects (KBOs) exhibit a broad range of colours, from completely neutral, to some of the reddest objects in the Solar System \citep[see for example][]{Tegler2003a}. These objects fall into at least two distinct classes based on their surface colours \citep{Tegler2003b,Barucci2005a,Fraser2012,Peixinho2012,DalleOre2013,Lacerda2014}. These two classes are found throughout the Kuiper Belt and centaur regions, with no significant detected differences in the distribution of colours between different dynamical classes \citep{Fraser2012,Peixinho2012,DalleOre2013}. This suggests that the bimodal colour distribution of KBOs and centaurs is not due to a current evolutionary process, but rather, a primordial property of the entire population, and likely a consequence of their compositions. 

Two competing types of classification exist, based on the distribution and clumpiness of the observed colours. \citet{Fraser2012} find that the small KBOs bifurcate into only two classes, with each class occupying a continuum of optical and NIR colours. \citet{Barucci2005a,DalleOre2013} suggest an alternative, that KBOs occupy a larger number of discrete classes with unique colours describing each class. Spectral variability holds the key to determining which of these classifications is correct. Consider an object which exhibits surface heterogeneity, and therefore, spectral variability. The case in which an object belonging to a certain class exhibits a surface consistent with another class must be quite rare. Otherwise, that object will often be observed with intermediate colours between the two classes. If this were common to KBOs, little clumping in the colour distribution would be observed, and the two separate classes would appear as one class of object in the global colour distribution. Clearly, spectral variability has the potential to distinguish between the continuum classification and more discrete schemes.

Over the last decade, astronomers have reported various detections of spectrally variable KBOs. Two detections of surface heterogeneity from spectro-photometry are presented by \citet{Choi2003} and \citet{Lacerda2008}. \citet{Lacerda2008} present the clear and confident identification of a red spot on the surface of KBO 136108, Haumea which manifests itself as a variation in the observed optical and NIR colours correlated with Haumea's variable light-curve. \citet{Choi2003} present VR photometry of object 1999 TD10. They find that TD10 exhibits wild optical colour variations between one of the reddest objects in the known Kuiper Belt, and having colours significantly bluer than Solar. This detection seems questionable as such a large variation in colour, if common to KBOs, should erase the bifurcated optical colour distribution that has so confidently detected. Luckily, 1999 TD10 belongs to the sample presented here.

Other reported heterogeneity comes from optical and NIR spectral observations of a few unique KBOs. \citet{Barucci2002} and later \citet{Merlin2005} and \citet{Licandro2005} all reported variability in the water-ice absorption features exhibited by the Centaur 32532, Thereus. \citet{deBergh2004,Fornasier2004} reported variations in the optical spectral slopes of Plutinos 47932 and 38628 - Huya. \citet{Fornasier2009} observed a lack of an absorption feature at $\sim0.7\mbox{ $\mu$m}$ in the optical spectrum of KBO 208996, 2003 AZ84, a feature detected in earlier spectra. Finally, \citet{Merlin2010} demonstrated that optical and NIR photometry of Centaur Echeclus does not agree with its observed spectrum. Detection of surface heterogeneities from optical and NIR spectra are exceptionally challenging, often plagued by subtle calibration issues. These issues are notoriously difficult to diagnose, and can show themselves as ``features'' in the resultant spectra \citep[see][for a discussion of the difficulties in calibrating ground-based spectra]{Bus2002}. Unfortunately, to-date, the detections of heterogeneity through spectra are all of a level similar to what may arise from calibration issues making their detection uncertain.

Here we present optical and NIR spectro-photometry from the Wide Field Camera 3 (WFC3) aboard the Hubble Space Telescope (HST) of 13 KBOs along with past photometry presented by \citet{Fraser2012} taken as part of the Hubble/WFC3 Test of Surfaces in the Outer Solar System (H/WTSOSS) and of target 2007 OR10 presented by \citet{Brown2011b}. Given the sensitivity stability of the HST (less than 1\% variation in zero points for the cycles during which our observations were taken) and the avoidance of any atmospheric extinction calibration difficulties, the photometry we present here represent the best evidence to date of surface heterogeneity on a moderate sample of KBOs. In Section 2 we present our observations, photometry, and resultant coarse spectra. In Section 3, we present a discussion of the detection of light curve variability on 8 of our targets, 4 of which also exhibit significant spectral variability, and conclude with a short discussion in section 4.

\section{Observations and Reductions}
Here we present new observations of 13 KBOs taken with the Wide-Field Camera 3 (WFC3) aboard the Hubble Space Telescope (HST) during cycle 18 as part of H/WTSOSS. The observations published in \citet{Fraser2012} were processed with version 2.7 of the CalWF3 processing pipeline. The pipeline has seen moderate improvement since, including slight changes to the photometric zero points, and identification of additional bad pixels. A reanalysis of those data using pipeline version 3.1 was warranted. An improved photometry routine compared to that used in \citet{Fraser2012} was used on both the cycle 17 and cycle 18 data, and is described as follows.

\subsection{Cycle 18 Observations}
The photometry we present here were observed as part of the cycle 18 under GO-Program 12234 as an extension to the H/WTSOSS program \citep{Fraser2012}. 13 KBOs were targeted by this program. Targets were selected from the original program to create a sample spanning the full range of dynamical class, brightness, absolute magnitude, and colour. The original purpose of this program was to gather spectral photometry in the wavelength regions between the filters observed in \citep{Fraser2012}. F606w was repeated during cycle 18 to ensure brightness normalization was possible between the separate cycle 17 and 18 visits. The original purpose of the program was to search for potential absorption features on faint KBOs. As will be demonstrated below however, the main result of these observations is the clear detection of spectral variability on a number of our targets.

All targets were observed and reduced in an identical manner. Each target was observed during a single 44 minute observability window, during which the telescope was slewed at the apparent rate of motion of the target. During the window of observation, two images where acquired in each of the F606w (a), F775w (b), F098m (c), and F101w (d) filters, and four images in F127m (e). The images were taken in an ab-abcdee-cdee pattern, with 2" dithers separating exposures, marked by the dashes. We utilized exposure times of 128~s, 114~s, 115~s, 207~s, and 275~s in the F606w, F814w, F110w, F098m, and F127m filters, respectively. This observing pattern was chosen to afford some resilience to the effects of cosmic rays, and the possibility of detecting rotational variability during a sequence, all while ensuring that the data acquired could be fully downloaded during Earth occultation.

We started with the non-multidrizzled, CalWF3 v. 3.1 processed images. Full details of the pipeline are available in the CalWF3 manual, and include bias and dark subtraction, flat field corrections, pixel noise model generation, and conversion to flux units for the IR detector. An accurate centroid of the target KBO in each image was first determined. This was done by moving a 10x oversampled theoretical TinyTim \citep{Krist1993} point spread function (PSF) in a 0.01 pixel spacing grid in x and y. The best-fit centroid was found as the coordinates which minimized the chi-squared between the PSF and the observed KBO image. As the coordinates used by TinyTim do not match the photo centre of the theoretical PSF, the source photocenter was measured in a 4 pixel radius aperture about the PSF image at the best-fit coordinates. Finally, photometry was measured from the KBO using a circular aperture, centred at the photocenter. This combination of steps was found to best minimize centroiding errors on faint sources. Aperture corrections were calculated from the theoretical PSF. Uncertainties were calculated by summing the pixels of the CalWF3 flux uncertainty extensions, in each aperture, using standard error propagation. 

For all targets other than 1998 SM165, the photometry aperture used had a 4 pixel radius in all filters, corresponding to 0.16" and 0.52" in the UV and IR detectors, respectively. 1998 SM165 is a known binary which is partially resolved in the UV channel (F606w and F775w filters) and unresolved in the IR channel. A 9 pixel radius aperture was used to gather the flux from both sources. In the remaining filters a 4 pixel radius aperture was used.

The magnitude in each filter was calculated as the mean of that filter's measurements. Visual inspection of the PSF radial profile after normalizing to the peak of each source was used to reject bad measurements; the profiles of each image pair were overlapped allowing easy identification of background sources and cosmic ray strikes which would affect the photometry. In addition, if the difference in magnitude between the two measurements of a pair was more than $3-\sigma$ discrepant, the brighter of the two was rejected, as the most likely cause of dramatic apparent brightness variations is cosmic ray hits and background sources, both of which would be rejected by this procedure. An example of a rejected measurement can be seen in Figure~\ref{fig:psfFig}. We report the cycle 18 photometry in the natural STMAG magnitude system, in Table~\ref{tab:cycle18}. As no phase curve  information is available for our targets, we report absolute H$_{\textrm{606}}$ magnitudes corrected for distance only. 

Object 2001 QX322 was found to be offset from its predicted position at the time of observation, and as such fell on the edge of the optical images. As a result, no optical photometry of 2001 QX322 are available from these observations.

\subsection{Cycle 17 Observations}
Using the techniques discussed above, we reanalyzed the cycle 17 photometry. For those sources with no obvious binary companions, 4 pixel radius apertures were used in the F606w and F814w filters, and 3.5 pixel radius apertures in the F139m and F153m filters. 

For known binaries 1998 SM165,  2001 QC298,  the secondaries were partially resolved in the optical images. For these objects, a single large aperture was used to gather the flux of both targets simultaneously. For known binary, 2003 HG57, a single aperture was used for both sources in the IR, and two small apertures of 4 pixel radii used in the optical, with appropriate aperture corrections applied. For binary 2001 QT297, the secondary was fully resolved in all images, and small apertures used accordingly. In all cases, we report the photometry of both sources combined. We will discuss resolved photometry in a separate manuscript.The reanalyzed photometry is presented in Table~\ref{tab:cycle17}. Differences between the photometry presented in \citet{Fraser2012} and the reanalysis presented here are due mainly to better bad pixel identification and slight zero point adjustments between versions 2.7 and 3.1 of the CalWF3 pipeline.

For 1998 SM165 we also consider the colour (F606w-F814w)=$1.12\pm0.03$ in the Vega magnitude system reported by \citet{Benecchi2011} in our analysis of that target's spectral variability.

\subsection{Photometric Consistency}
The photometry we measure between the programs has proven to be consistent and repeatable for those objects which exhibit no significant lightcurves. This is demonstrated by objects 2007 OR10 and 2004 EW95 which have no more  than $0.01$ magnitude difference in F606w magnitude between separate visits of each object. This difference is consistent with the expected photometric precision, and is a result of WFC3's stable photometric sensitivity and our photometric routine.

The WFC3 photometric zeropoints have proven to be exceptionally stable over the time period in which our observations were taken. \citet{Kalirai2011} find less than $\sim1\%$ variance of the zeropoints over 1.5 years of observations of four photometric standards.

\begin{figure}[h]
   \centering
   \includegraphics[width=6.5in]{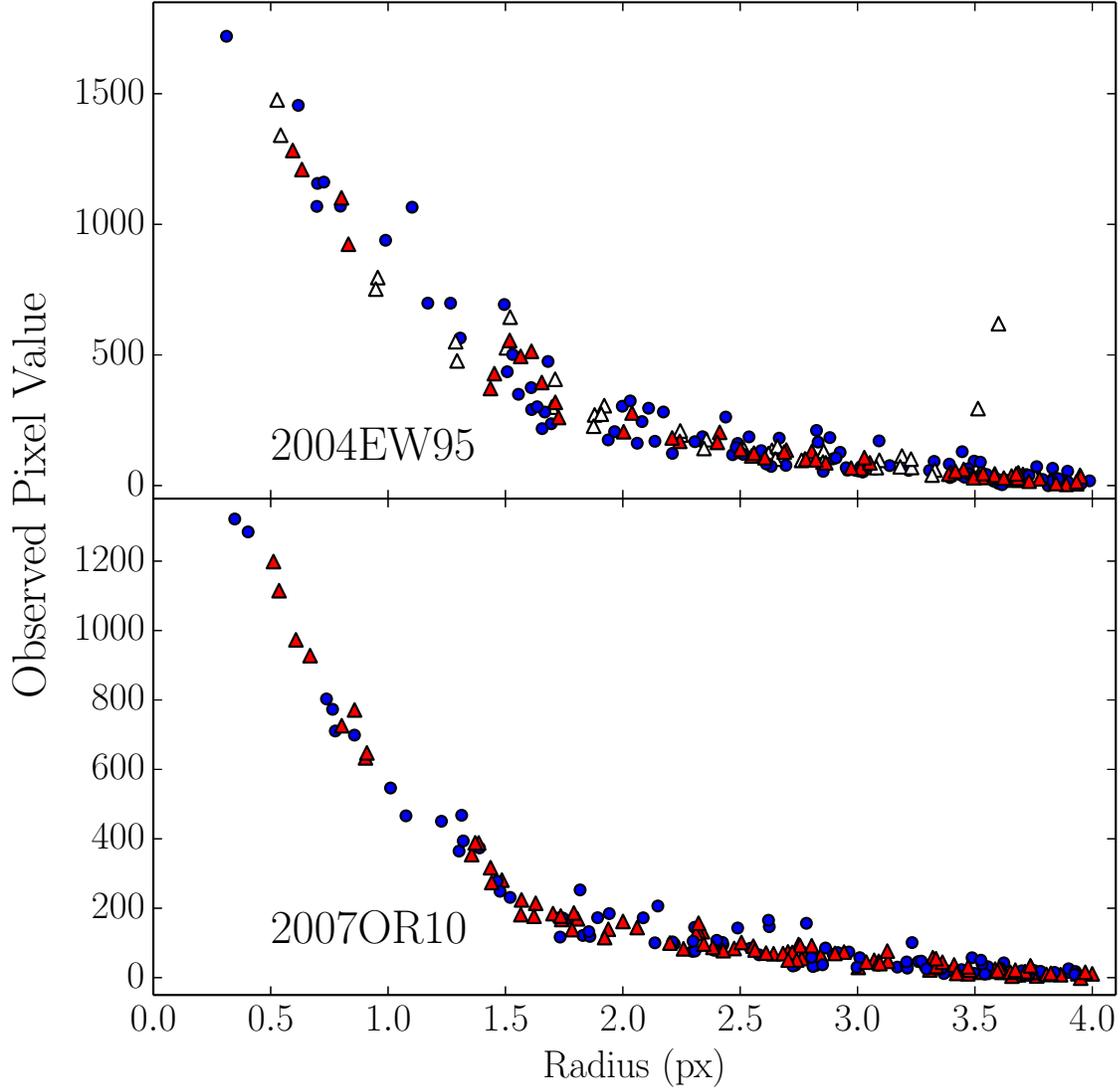}
   \figcaption{Measured radial Point Spread Functions (PSFs) of 2004 EW95 (top) and 2007 OR10 (bottom) in the F606w filter for observations taken in cycle 17 and 18, shown as blue circles and red triangles respectively. One image of 2004 EW95, in cycle 18 was affected by a cosmic ray strike within the photometric aperture. The PSFs of that image is shown by the white points, and the photometry from that image was rejected by our routine. The cosmic ray strike can be seen as higher than usual points in the white triangles at  $\sim3.5$ pixels radius. \label{fig:psfFig}}
\end{figure}

Our procedure of first a visual rejection, then by magnitude difference proved excellent at rejecting those photometric measurements affected by cosmic rays or hot pixels.  We demonstrate this with objects 2007 OR10 and 2004 EW95 in Figure~\ref{fig:psfFig}. 

For object 2007 OR10, background sources or cosmic ray strikes did not affect any of the four images taken in the F606w filter. It can be seen that the radial PSFs of all four images are very similar. All four measurements have a scatter about the mean of less than 0.01 mags, consistent with variation due to background noise, and were accepted by our photometry routine. This was not the case for 2004 EW95. Of the four images taken in the F606w filter, one was affected by either cosmic ray strikes or background sources within the photometric aperture. The radial profile allowed easy identification of large deviations of the source PSF away from those not affected by these issues (see the white points in Figure~\ref{fig:psfFig}). That image was rejected by our photometry routine. Photometry of the remaining two images differ by less than 0.04 magnitudes, consistent with expectations due to Poisson noise. Repeat visits to each of these non-variable objects demonstrates the absolute consistency of our photometry from visit to visit.

\subsection{Ground Based Observations}
BVRIJHK photometry of 1998 SM165 are reported in \citet{Delsanti2004}. In addition, we utilized photometry of 1999 RB216, 1999 TD10, and 2001 QX322 reported in the MBOSS database\footnote{\url{http://www.eso.org/~ohainaut/MBOSS/MBOSS2.dat'}} \citep{Hainaut2002}. Colours of 2004 EW95 have been previously reported in \citet{Mommert2012} (V=$21.04\pm0.3$, (B-V)=$0.70\pm0.02$, (V-R)=$0.38\pm0.02$) and \citet{Perna2013} (V=$21.03\pm0.03$, (B-V)=$0.68\pm0.04$, (V-R)=$0.38\pm0.04$, (R-I)=$0.52\pm0.05$). 

Unpublished observations of 2004 EW95 were identified with the Solar System Object Image Search tool\footnote{\url{http://www3.cadc-ccda.hia-iha.nrc-cnrc.gc.ca/en/ssois/index.html}} provided by the Canadian Astronomy Data Centre \citep{Gwyn2012} taken at the Very Large Telescope with the FORS2 instrument during the 2nd and 3rd of April, 2011 in support of the TNOs are Cool program \citep{Perna2013}. The standard calibration images were saturated. Zeropoint calibration was possible using ugriz photometry of background stars observed in the Sloan Digital Sky Survey with conversion to the BVRI systems\footnote{\url{http://www.sdss.org/dr7/algorithms/sdssUBVRITransform.html}}. Photometry was extracted using standard aperture photometry techniques, and was ultimately limited by the zero point calibrations. 2004 EW95 was found to have a visual magnitude V=$20.86\pm0.01$, and colours (B-V)=$0.74\pm0.03$, (V-R)=$0.38\pm0.04$, and (R-I)=$0.41\pm0.07$. Here we adopt the mean of the observed colours (B-V)=$0.70\pm0.03$, (V-R)=$0.38\pm0.03$, and (R-I)=$0.46\pm0.06$.

\section{Results \label{sec:results}}
Here we present the results of our cycle 17 and cycle 18 photometry including a reanalysis of the colour classes, and the detection of significant light curve and spectral variations on many of our targets.

\subsection{Compositional Classes}
\begin{figure}[h]
   \centering
   \includegraphics[width=6.5in]{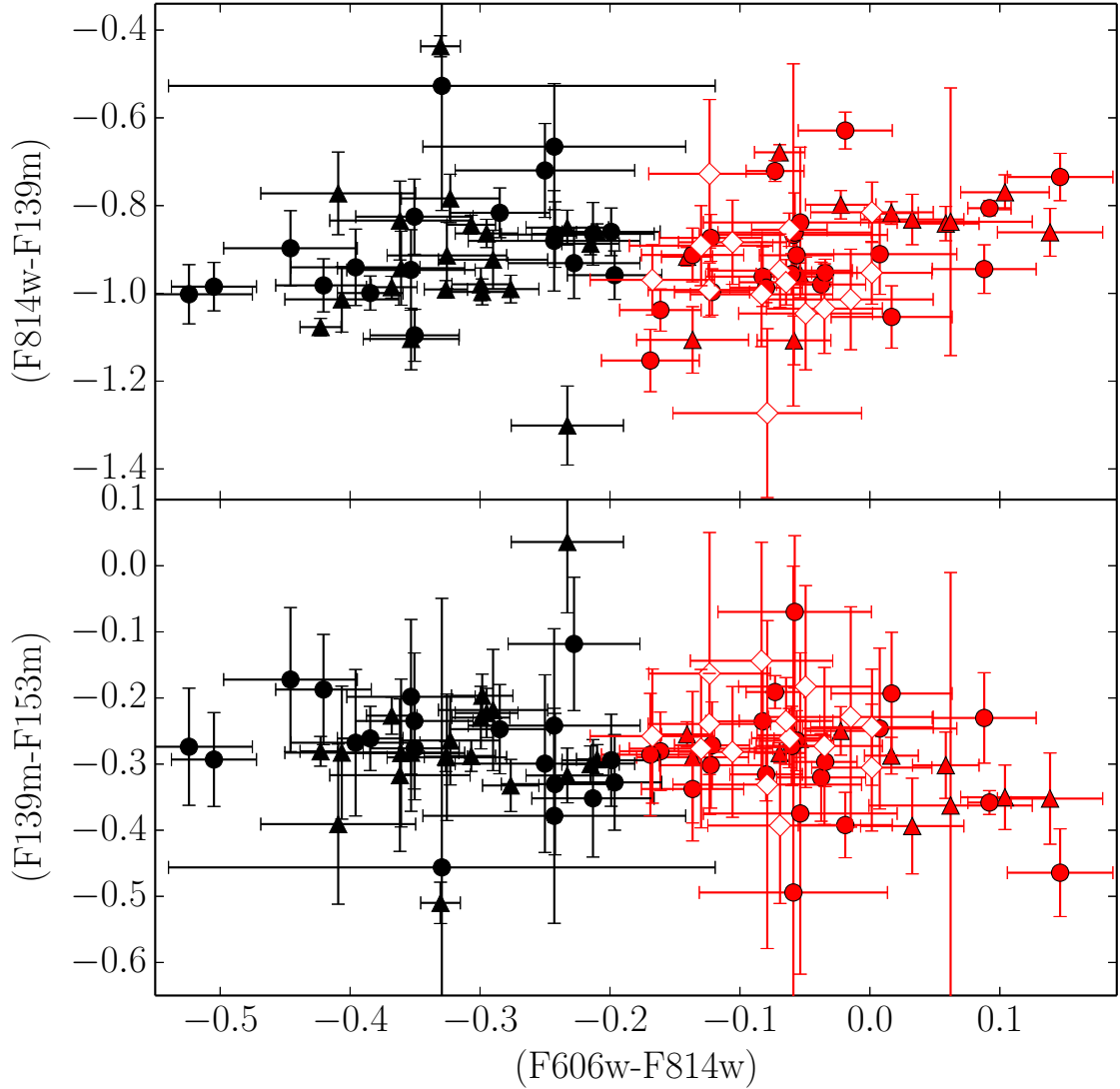}
   \figcaption{Optical and infrared colours of all cycle 17 targets with $H_{\textrm{606}}>6$. Colours correspond to the neutral (black points) and red (red points) classes identified by the MST clustering. Triangles, circles, and open diamonds represent the excited objects with $q<30$~AU, $q>30$~AU, and the cold classical objects, respectively. \label{fig:ccc}}
\end{figure}

In Figure~\ref{fig:ccc} we present the colours of the cycle 17 targets with $H_{\textrm{606}}>6$. We apply the same minimum spanning tree (MST) and F-statistic based clustering technique, and the FOP test to the sample as that utilized in \citet{Fraser2012}. This routine searches for clusters in the dataset by pruning a branch of the Euclidean minimum spanning tree, dividing the sample into two clusters or subsets. The branch that is chosen is that which maximizes the F statistic defined as 
\begin{equation}
F=2\left(\frac{min(|t_1|,|t_2|)}{N}\right)\left(\frac{l}{l_{\textrm{max}}}\right)
\end{equation}
\noindent
where $t_1$ and $t_2$ are the number of objects in each of the two subpopulations, $N=t_1+t_2$ and $l$ and $l_{\textrm{max}}$ are the length of the pruned branch and longest branch in the tree, respectively.  This method generally selects two subpopulations which are maximally separated in a Euclidian sense. The FOP test then determines the probability of finding a value of F as large as observed from an equal sized sample with similar measurement uncertainty distribution. This is done by creating random samples bootstrapped from the observed colours and uncertainties.

The colour distribution bifurcates at (F606w-F814w)$=0.19$ into the red and neutral classes of object, in agreement with the observed bifurcation of optical colours of the centaurs and dynamically excited objects. The FOP test identifies the bifurcation at greater the 70\% significant while Hartigan's DIP test finds only a 3\% probability that the total colour sample of small objects is unimodal. 

Our reanalysis repeats the same findings as we previously found, and can be summarized as follows:

\begin{itemize}
\item The dynamically excited KBOs bifurcate into two colour classes based on their optical colour.
\item Each class exhibits correlated optical and infrared colours. The Spearman rank test gives probabilities of 4\% and 0.6\% that the neutral and red classes do not exhibit correlated colours.
\item The centaurs ($q<30$~AU) and more distant excited objects do not exhibit different colour distributions. What ever apparent differences exist cannot be distinguished from observational biases that affect the sample, including the known trend of visible albedo and colour \citep{Stansberry2008,Fraser2012,Lacerda2014}.
\item The cold classical objects, with $41<a<45$~AU and $i<5^\textrm{o}$ fall on the neutral end of the red class of object. All cold classical objects we observe belong to the red class of object.
\end{itemize}

The recent work of \citet{DalleOre2013} suggests that the next division based on optical and NIR colours would divide the neutral class at (F606w-F814w)$\sim-0.35$ and (F814w-F139m)$\sim-0.9$. We find no evidence in support of this additional level of complication. On the contrary, as presented below, the colours of 2005 TV189 suggest that the neutral class we identify here is not a sum of overlapping smaller classes, but rather is one continuous class.

\subsection{Brightness Variations}
\begin{figure}[h]
   \centering
   \includegraphics[width=6.5in]{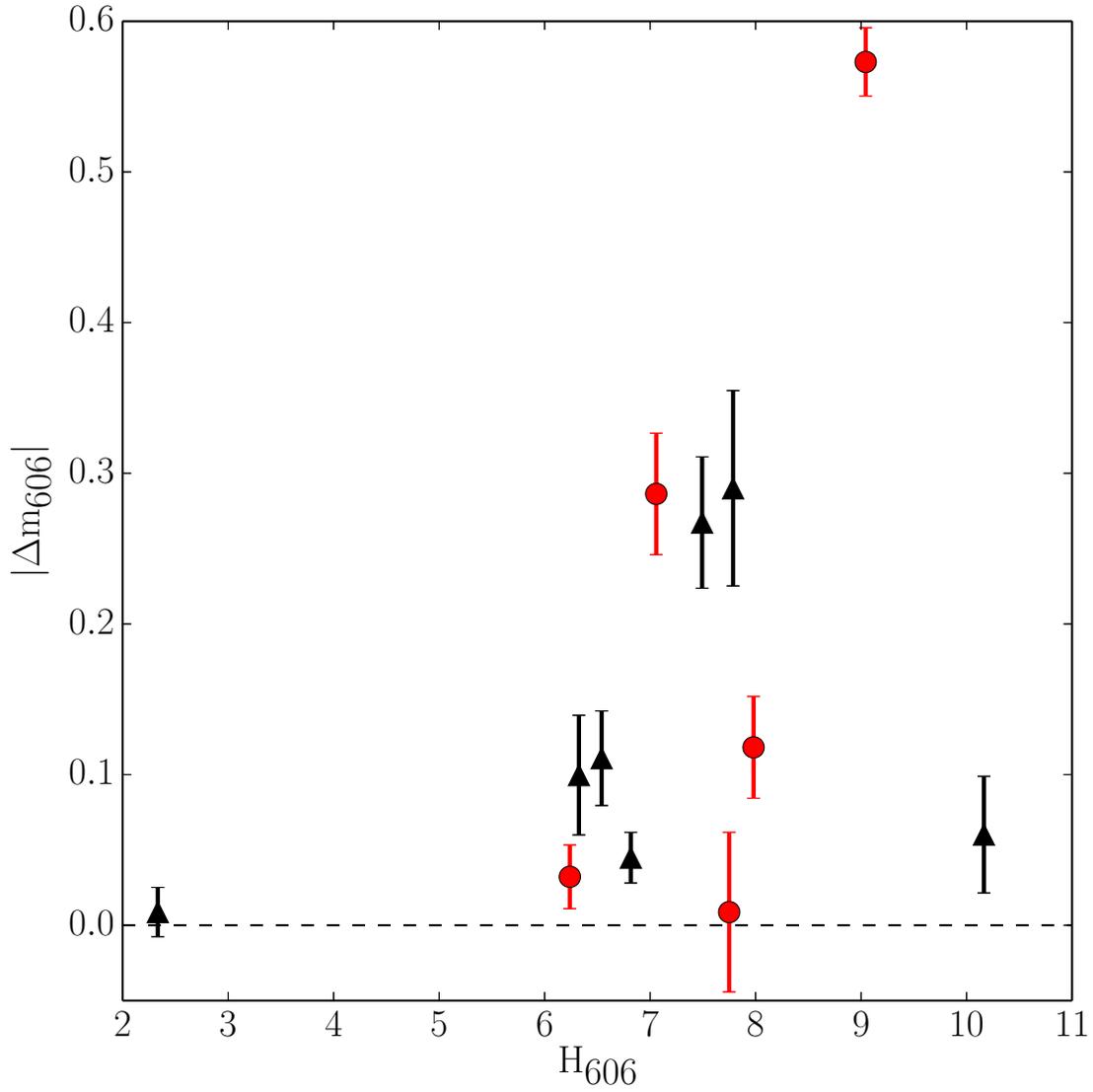}
   \figcaption{Absolute magnitude of the difference in F606w repeat measurements from cycle 17 and cycle 18 versus the F606w absolute magnitude of the targets. Red circles and black triangles represent those objects that do and do not exhibit spectral variability respectively (see Section~\ref{sec:specVar}. \label{fig:deltaH}}
\end{figure}

Comparison of the dual epochs of F606w photometry, it is clear that most of our targets exhibit significant brightness variations. We demonstrate this in Figure~\ref{fig:deltaH} where we plot $|\Delta m_{\textrm{606}}|$, the absolute value of the difference in measured F606w magnitude between cycles 17 and 18. 6 of the 12 objects with two F606w measurements exhibit variations of an amplitude greater than the 3-sigma uncertainty on the measurement difference. The variations of the measured F606w magnitude between the cycle 17 and cycle 18 visits for those 6 objects range between 0.1 and 0.6 magnitudes.  We note here that ground-based observations detect a $\sim0.1$ magnitude r'-band variation in the brightness of 2004 EW95 that we do not detect here. The consistency of EW95's F606w photometry between the two visits may just be chance, and this object may exhibit significant variability.

It is possible that some of the small magnitude variations are due to varying phase angle of the targets between observations. Little information about the phase curves of our targets is available. We can however, estimate the effects of phase using typical behaviour of the few KBOs for which phase information is available; we adopt a linear phase curve with slope $\beta=0.14 \mbox{ mag deg$^{-1}$}$ as done in \citet{Perna2013}. Comparing the estimated brightness variations to those observed, we see that only the variations of 2005 TV189 can be accounted for. All other observed variations are too large  or posses the incorrect sign to be consistent with phase variations exhibited by other KBOs.

Of the targets with $H_{606}>6$, 6 out of 11 present significant brightness variations inconsistent with phase effects. Interestingly, assuming sinusoidal light curves, excluding the large object 2007 OR10 which exhibits no detectable brightness variations, our entire sample is fully consistent with all objects having light curve amplitudes at least as large as 0.26 mag, the mean amplitude of all measured light curves in the Kuiper Belt found by \citet{Duffard2009}. That is, our sample is consistent with all small objects having significant light curves. This sample chosen for observations during cycle 18 was selected for their observed cycle 17 colours and not their variability. This suggests that the majority of KBOs smaller than $H_{606}\sim6$ exhibit light curve variations of greater than a couple tenths of a magnitude.

\subsection{Spectral Variations \label{sec:specVar}}

\begin{figure}[h]
   \centering
   \includegraphics[width=3.5in]{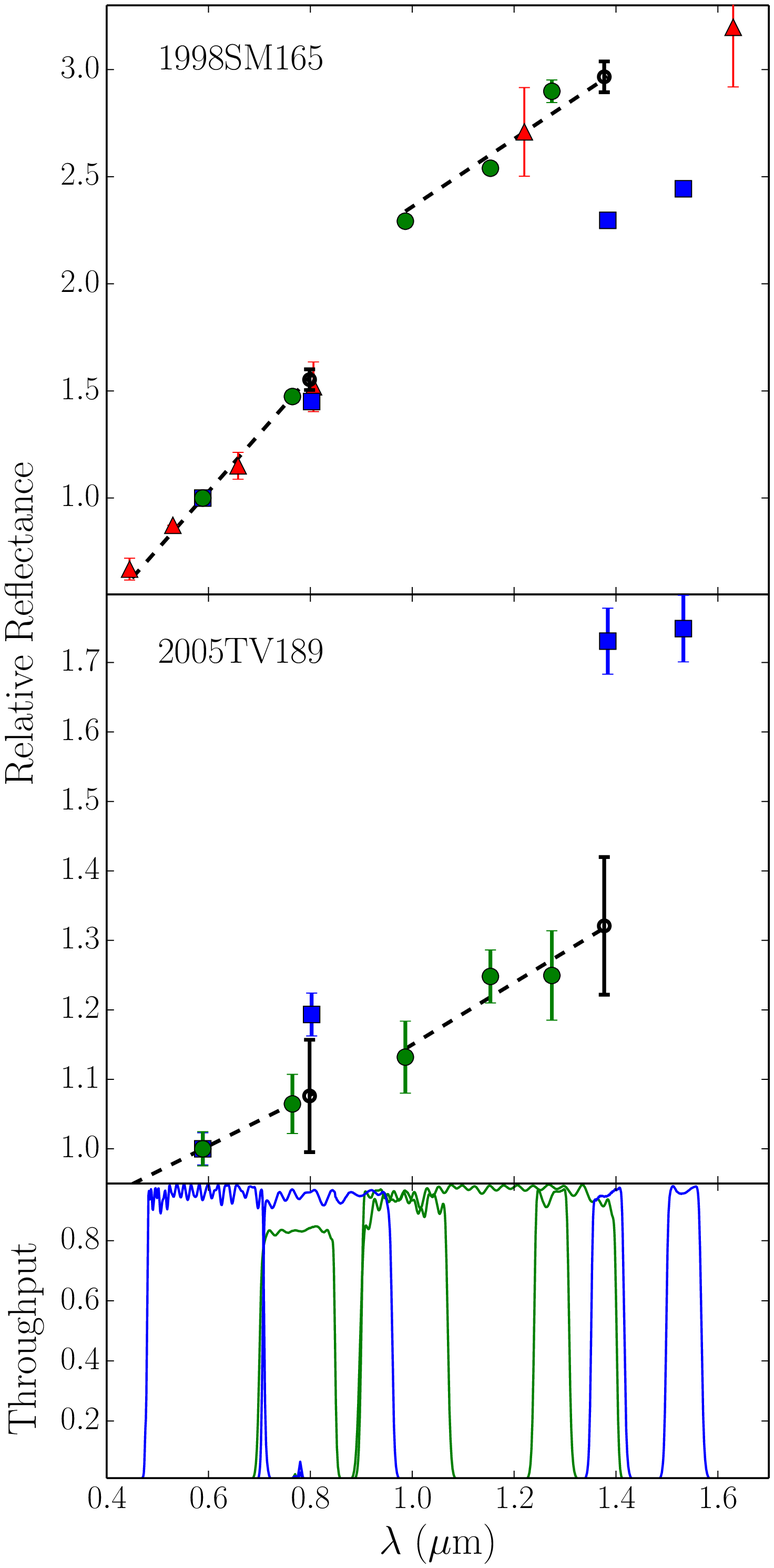}
   \figcaption{ Reflectance spectra of the spectrally variable objects 1998 SM165 and 2005 TV189. Spectra have been normalized to  F606w filter. Blue squares and green circles mark the spectra derived from the cycle 17 and cycle 18 photometry respectively. Black open circles represent the estimated F814w and F139m photometry from the linear regression of the cycle 18 optical (F606w and F775w) and NIR (F098m, F110w, F127m) photometry, which are shown by the dotted lines. The filter bandpasses used in the cycle 17 and 18 photometry are shown by the blue and green curves respectively in the bottom panel. Red triangles are spectra derived from ground-based observations. \label{fig:spec1}}
\end{figure}

\begin{figure}[h]
   \centering
   \includegraphics[width=3.5in]{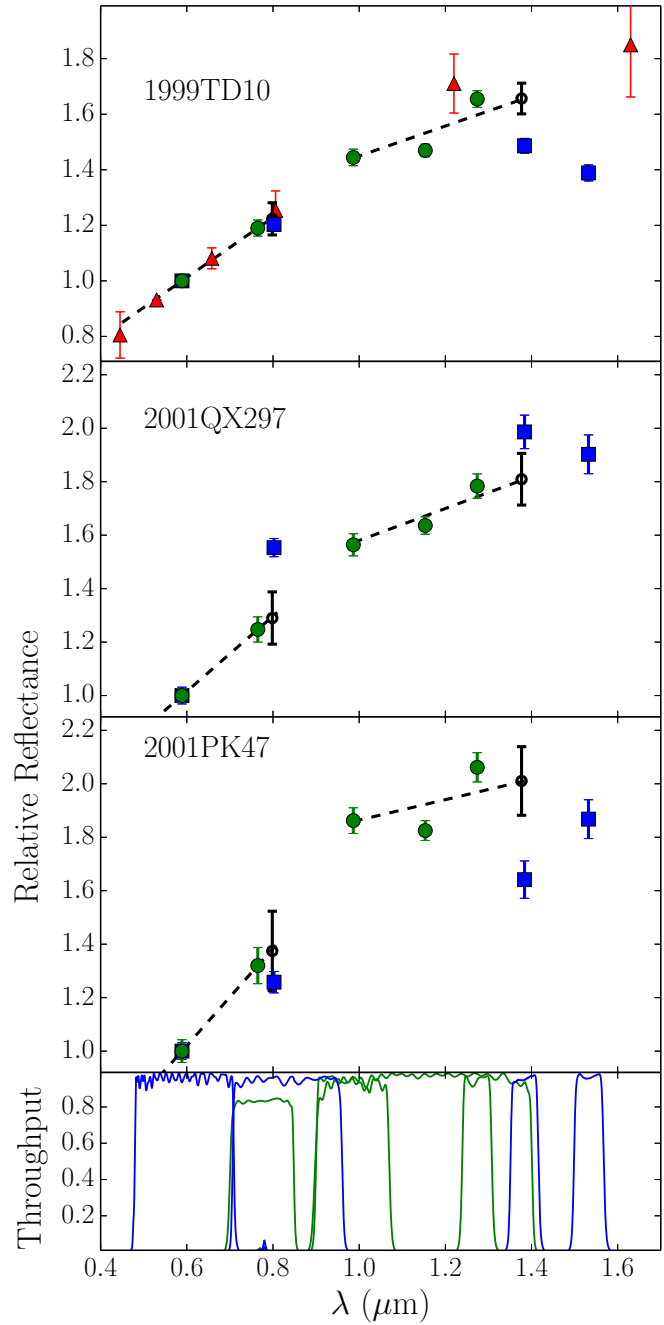}
   \figcaption{As in Figure~\ref{fig:spec1}, but for objects which show 2-sigma spectral variations. \label{fig:spec2}}
\end{figure}

\begin{figure}[h]
   \centering
   \includegraphics[width=3.5in]{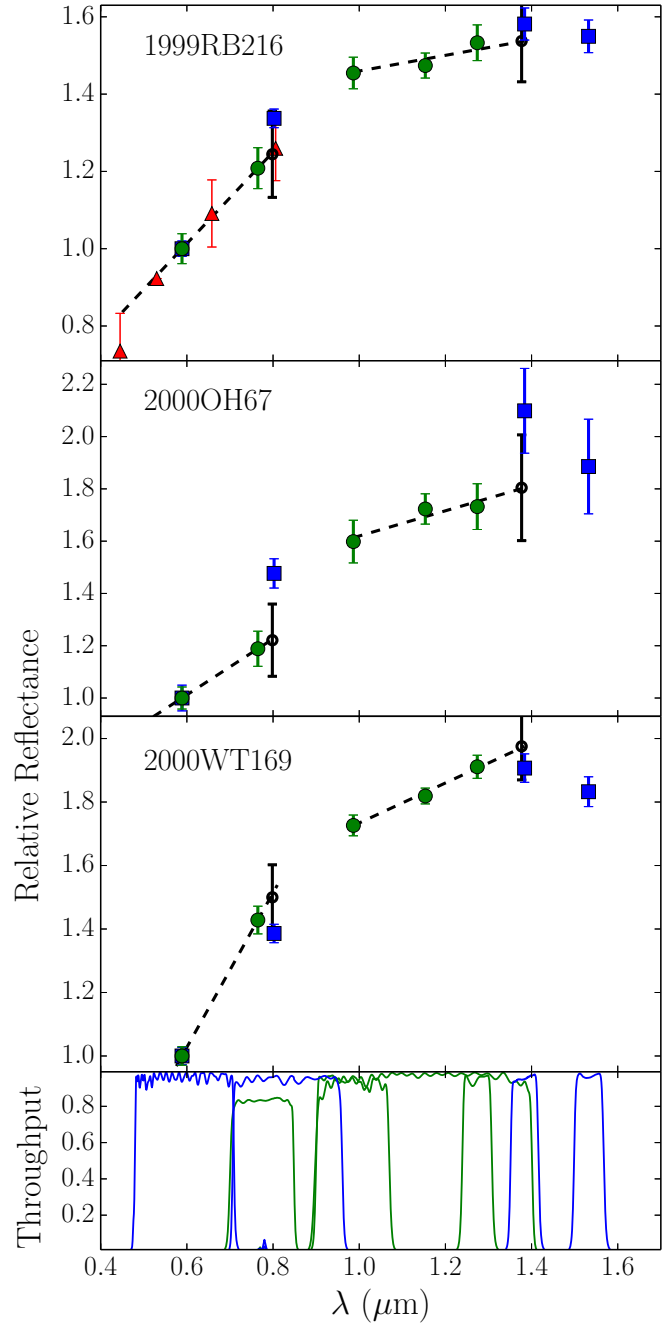}
   \figcaption{As in Figure~\ref{fig:spec1}, but for objects which show no spectral variations. \label{fig:spec3}}
\end{figure}

\begin{figure}[h]
   \centering
   \includegraphics[width=3.5in]{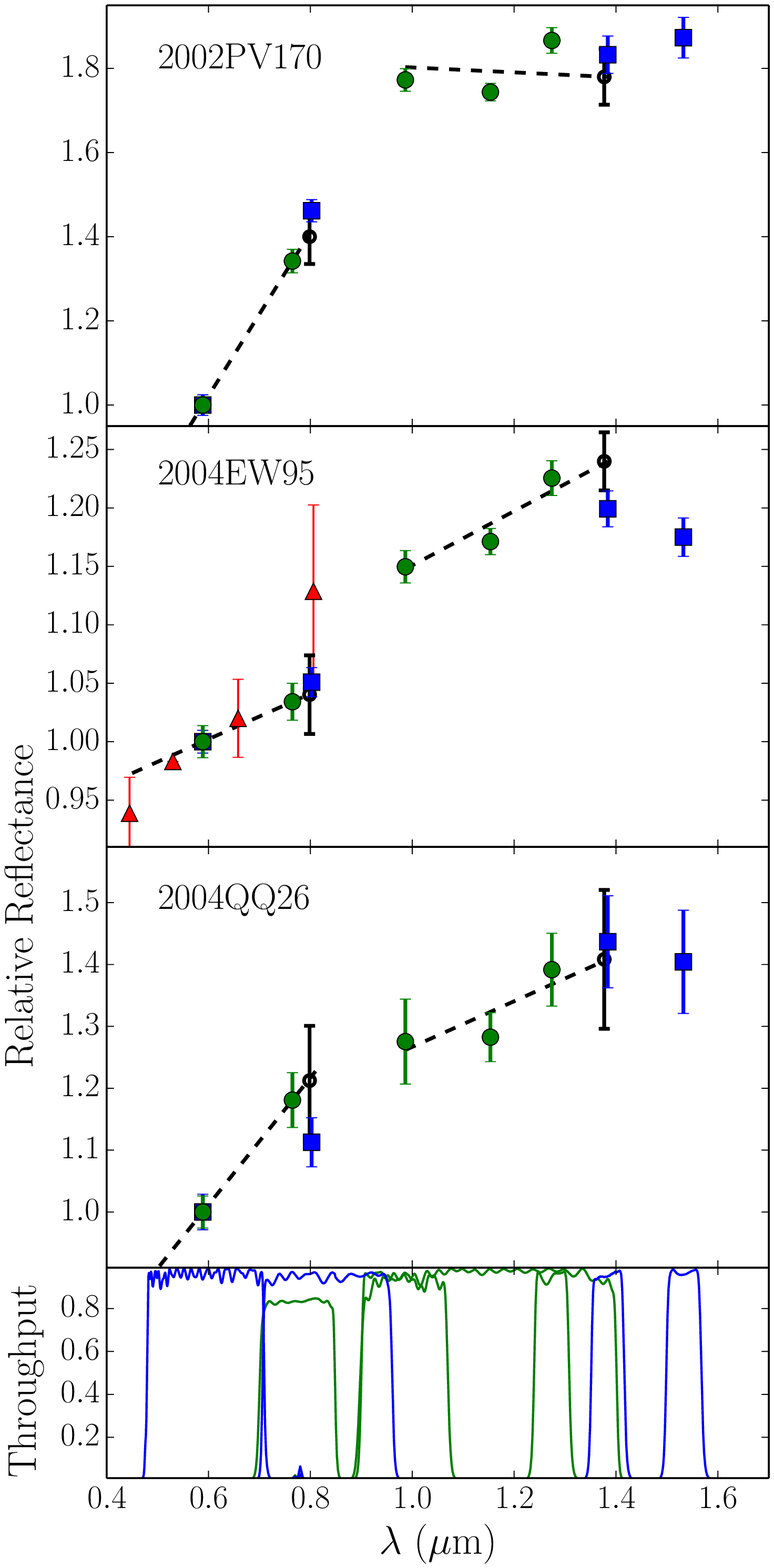}
   \figcaption{As in Figure~\ref{fig:spec3}.  \label{fig:spec4}}
\end{figure}

\begin{figure}[h]
   \centering
   \includegraphics[width=3.5in]{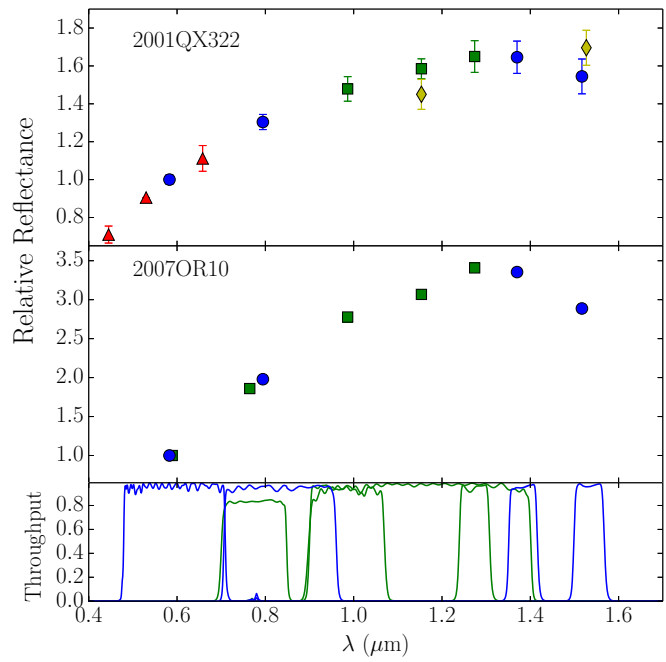}
   \figcaption{As in Figure~\ref{fig:spec3}.  The yellow diamonds mark the spectra obtained for 2001 QX322 from the data presented by \citet{Benecchi2011}.  The F098m, F110w, and F127m photometry of 2001 QX322 have been vertically scaled for visible representation to match the F139m photometry and the photometry of Benecchi.   \label{fig:spec5}}
\end{figure}

Relative reflectance spectra were obtained from the photometry by removing Solar colours from the observed spectra. Solar colours were determined from the {\it synphot} routine using Buser-Kurucz stellar analogs. We do not account for an uncertainty of $\sim0.01$ magnitudes as a result of not having a perfect Solar analog. Our reflectance spectra, normalized to 1 in the F606w filter are shown in Figures~\ref{fig:spec1} -- \ref{fig:spec5}. For object 2001 QX322, we also present spectra determined from the (F110w-F160w) colour presented by \citet{Benecchi2011}.

It is clear from Figures~\ref{fig:spec3} -- \ref{fig:spec4} that more than half of our targets do not exhibit spectral variations. That is, a smooth, continuous, approximately linear spectrum could simultaneously account for the spectra derived from the cycle 17 and cycle 18, and where available, ground-based photometry. For objects 1998 SM165, 1999 TD10, 2001 PK47, 2001 QX297, and 2005 TV189, this is not the case. We plot the spectra of those objects  in Figures~\ref{fig:spec1} and \ref{fig:spec2}.

These five objects appear to exhibit spectral variability. That is, the spectra derived from the cycle 17 photometry is inconsistent with that derived from the cycle 18 photometry. In particular, for SM165, TV189, TD10, and PK47, extrapolation of the  cycle 18  $\sim1 \mbox{ $\mu$m}$ photometry to longer wavelengths appears inconsistent with the cycle 17 F139m measurement. Extrapolation from the cycle 18 F775w filter to the cycle 17 measurement in the F814w filter reveals a similar inconsistency for 2001 QX297.

\begin{figure}[h]
   \centering
   \includegraphics[width=6.5in]{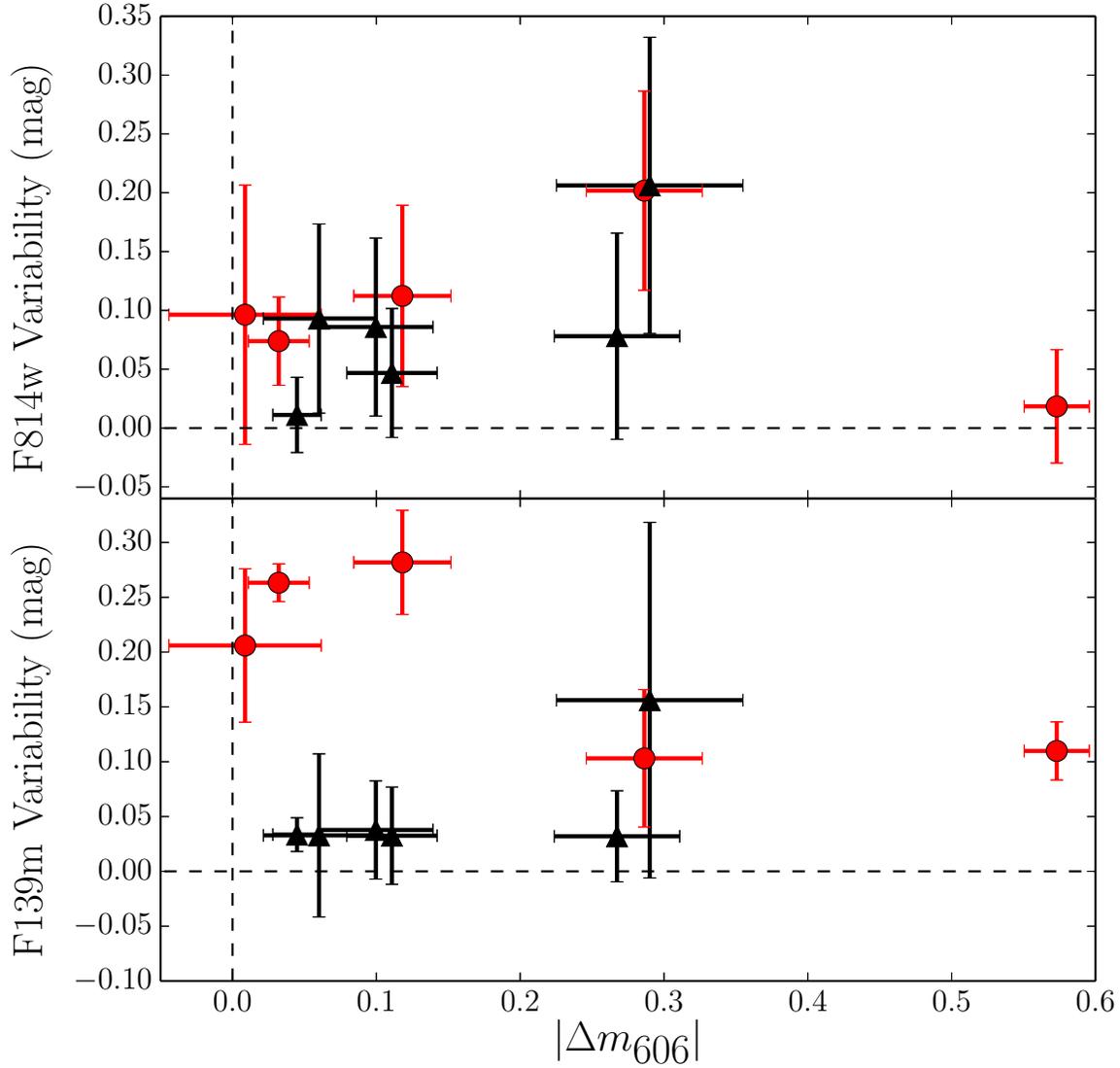}
   \figcaption{Spectral variability versus absolute magnitude of the difference in F606w repeat measurements from cycle 17 and cycle 18. \textbf{Top:} variability in the F814w filter. \textbf{Bottom:} variability in the F139m filter. Red circles and black triangles represent those objects that do and do not exhibit spectral variability respectively.\label{fig:variability}}
\end{figure}

We quantify these discrepancies in two ways. A linear fit to the F098m, F110w, and F127m filters from cycle 18 was used to predict the magnitude in the F139m filter. Similarly, a linear fit to the cycle 18 F606w and F775w filters was used to predict the magnitude in the F814w filter. We plot the difference between predicted and measured magnitudes from cycle 17 in Figure~\ref{fig:variability}. As can be seen, 1998 SM165, and 2005 TV189 all exhibit a difference between predicted and observed F139m magnitudes that is larger than the 3-$\sigma$ uncertainty (considering both observation uncertainty and uncertainty due to the linear fit). 1999 TD10 and 2001 PK47 exhibit greater than 2-sigma variations. Only object 2001 QX297 exhibits a significant difference between predicted and observed F814w magnitudes. Note: we exclude 2007 OR10 from this analysis as it is appears to have a spectrum affected by methane absorption, and thus, does not exhibit a linear spectrum in the NIR range \citep{Brown2011b}.

There are various phenomena that may manifest themselves as the spectral variations we detect. Some of our targets may exhibit large and rapid variations which may affect the photometry in different filters, and thus appear as spectral variations. We searched for such affects in our observations, by removing the colour average observed colours from the non-F606w photometry to provide estimated F606w photometry throughout the visit to a source. This was then plotted against time to search for temporal trends. Some hints of brightness variations through an orbit were detected, but not of an amplitude sufficient to account for the 20-50\% spectral variations we report here.

It may be the case that some previously undetected very sharp and very deep absorption features are causing the discrepancies we observe. While we cannot formally eliminate the existence of deep and sharp features as the source of the apparent spectral variability on these objects, that possibility seems unlikely, as the features required would be so sharp and deep as to be incompatible with typical solid state absorption features. The apparent spectral variability is also not due to sensitivity issues. As discussed above, the WFC3 filter zeropoints varied by less than 1\% across the cycles during which the data we utilize were taken \citep{Kalirai2011}. As such, calibrations cannot be the cause of the apparent spectral variation. It appears that these three objects truly exhibit spectral variability.

We should note that our detected spectral variability of 1999 TD10 is consistent in theme with the findings of \citet{Choi2003} who found that TD10 exhibits significant optical colour variations with amplitude large enough to be apparent in our data. Our observations do not corroborate their results however, as we find no significant optical variations. Like \citet{Choi2003}  we find that 1999 TD10 exhibits significant light curve variation of $\sim0.6$ mag. Drawing from their measured light curves, our observations happen to fall near a peak and trough of TD10's variability, and as such, should exhibit a $\sim0.3$ magnitude variation in $(F606w-F814w)$ between the two visits. This variation would be easily detected with the precision of the measurements we present here, and as such, our results are at odds with those of \citet{Choi2003}. It may be that their variations are entirely due to undetected cometary activity, which could explain the bluer-than-Solar colours they observed. The cause of this discrepancy however, remains unknown.

\section{Discussion}
The objects in our sample which exhibit statistically significant light curves exhibit a broad range of variability with $0.1\lesssim |\Delta m_{\textrm{606}}| \lesssim 0.6$. Our sample of objects with $H_{\textrm{606}}>6$ are all consistent with having variability with amplitude greater than 0.26 magnitudes, the mean of the light curve amplitudes found by \citet{Duffard2009}. Not all of the objects with detected brightness variations have associated spectral variations. In the Kuiper Belt, the reddest objects have higher albedos than neutral coloured objects \citep{Stansberry2008,Fraser2012,Lacerda2014}. Assuming the materials responsible for the observed spectral variations are not unique to those objects but is the same material found on other KBOs, then it follows that the brightness variations \textbf{all TNOs} are primarily shape driven, and not a result of compositional variations. Our observations suggest that the majority of objects with $H_{\textrm{606}}>6$ posses highly non-spherical shapes. We caution however, that the target list was chosen for their colour properties, and do not exhibit an unbiased sample. Therefore, our statements on the variability as a function of $H$ should be taken with appropriate caution.

Of our sample, 5 exhibit spectral variations. These 5 objects occupy a broad range of properties. They span essentially the entire range of dynamical classes: 1998 SM165 belongs to the scattered disk, 2005 TV189 is in a mean-motion resonance with Neptune,  2001 PK47 is a hot classical object, 1999 TD10 is a centaur (a scattered disk object in the system of \citet{Gladman2008}), and 2001 QX297 is a cold classical as we defined them in \citet{Fraser2012}. They span a factor of $\sim4$ in size, having absolute magnitudes $6\leq H_{\textrm{606}}\leq 9$. They occupy the full range of KBO colours from near Solar coloured to extremely red. Finally, they exhibit a broad range of light curve amplitudes. 2001 PK47 does not exhibit a significant light curve, while 1999 TD10 exhibits a large $\sim0.3$ magnitude amplitude. Interestingly, those objects which do not exhibit spectral variability occupy the same range of properties.  This suggests that what ever the source of the spectral variation, it cannot be driven purely by shape, size, or thermal environment. 

\begin{figure}[h]
   \centering
   \includegraphics[width=6.5in]{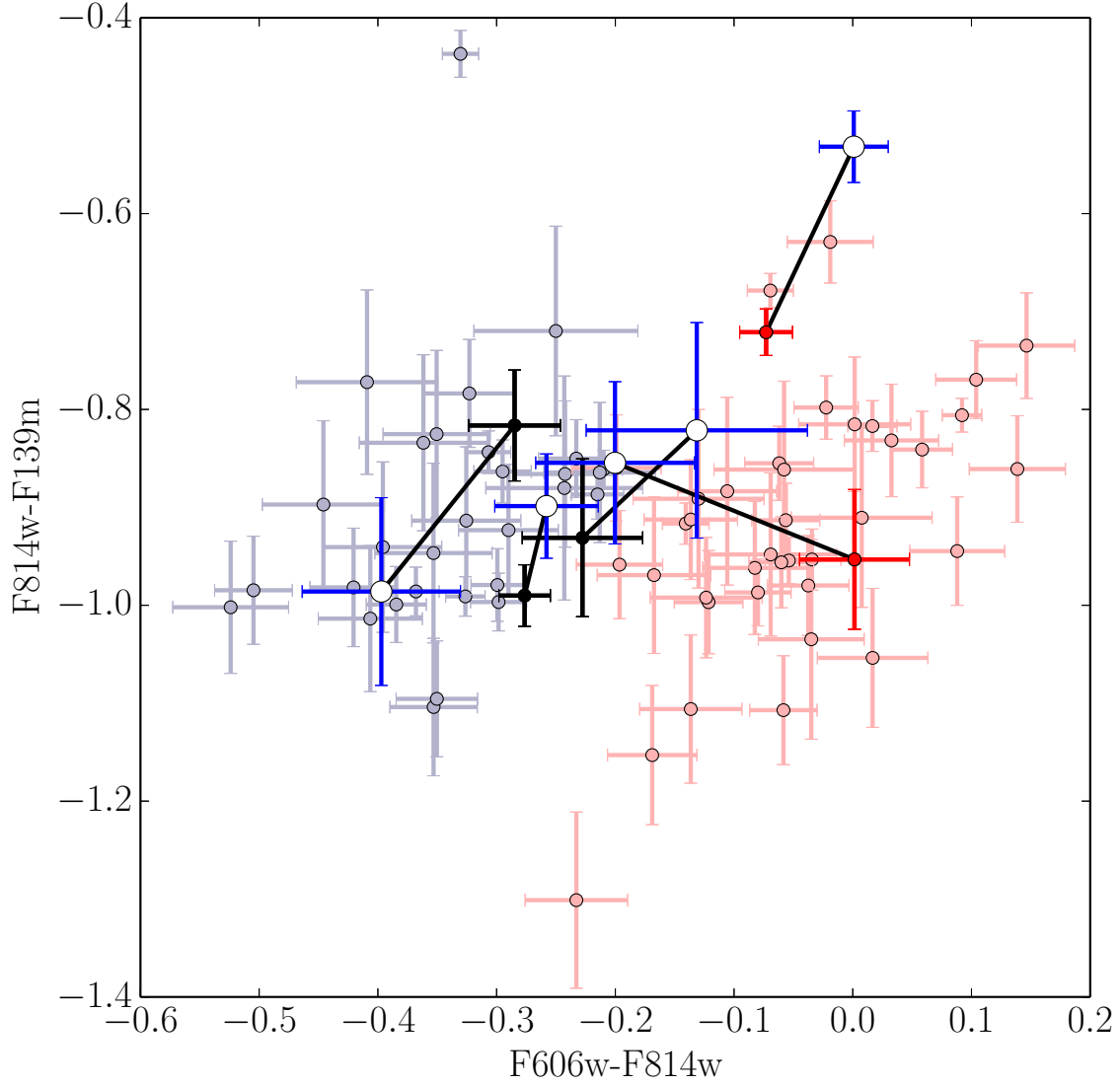}
   \figcaption{Observed and model optical and infrared colours of the 5 spectrally variable objects. The model cycle 18 colours are shown in blue points, while the observed cycle 17 colours are shown as solid red or black points according to their cycle 17 classification. Black lines connect the observed and model colours for each source. Gray and light red points show the full cycle 17 colours for comparison. \label{fig:modelCC}}
\end{figure}

One clue to the nature of the spectral variability and the colour distribution comes from the (F606w-F814w) and (F814w-F139m) variability. We present this in Figure~\ref{fig:modelCC}. From its cycle 17 and model photometry, 2001 PK47 has an ambiguous classification, having optical colour compatible with either red or neutral classes. 2001 QX297, a cold classical, and hence belonging to the red class, also has ambiguous model colour classification, with model colours sufficiently uncertain to (F606w-F814w)=-0.19 the colour division between the red and neutral classes. Objects 1998 SM165, 1999 TD10, and 2005 TV189 all have sufficiently accurate colours for classification. What is immediately apparent from Figure~\ref{fig:modelCC} is that the classifications for those three objects are the same between the two visits. That is, no object observed as red in cycle 17 was found to be neutral in cycle 18, and vice versa. 

Equally interesting as the consistency of classification, the difference in model colours for TV189 and TD10 cause each to span a broad range in both optical and infrared colours. TV189 in particular spans nearly the full range occupied by the neutral class. This argues strongly in favour of the neutral class being one continuous class rather than the overlap of underlying, more compact classes. The reasoning is simple; it must be that at times, TV189 exhibits colours intermediate between the two extremes we have observed. Thus, the neutral class, of which TV189 is a member, spans at least the range between the two observed colours. 

Another way to envision this is to consider the scenario in which the neutral class ((F606w-F814w)$<-0.19$) is made up of two populations, each possessing different optical and NIR colours. Further, consider the possibility that TV189 is the product of a merger of one object of each of those populations. It must be then, that these mergers occur frequently enough such that one of the only two spectrally variable neutral objects we observe is the product of such a merger. In this scenario, the two populations would be mixed together by these mergers, and no longer be well defined in a colour-colour plot. Rather, they would appear as one broad class, with some redder objects, some more neutral objects, and some having intermediate colours between the two. This is exactly the scenario we put forth in \citet{Fraser2012} to describe the compositional classes - a classes members are mixtures of different materials. We have insufficient data to draw similar conclusions for the red sample. It appears however, that the neutral class, is one class spanning a broad range of colours, rather than multiple discrete classes occupying smaller colour ranges.

As discussed above, it seems the classes cannot be due to thermal environment. Further, processes such as cometary activity or collisions, if dominant, would likely homogenize the colours of a class's members. As such, large variations like that of TV189 would not persist. Collisions can further be excluded as the source of the variability when one considers the collisional environment. \citet{Dell'Oro2013} has demonstrated that collisions in the Kuiper Belt typically occur at $1-2\sim\mbox{ km s$^{-1}$}$, meaning that most collisions are destructive, rather than constructive, a requirement to deposit a large amount of differently coloured material on an object. 

The most obvious conclusion from this result is that, at least the members of the neutral class possess a broad range of compositions. This is a natural outcome of the mixing model put fourth by \citet{Fraser2012}, in which the colours of a class are mainly determined by a mixture of materials unique to that class, under the assertion that some objects, like 2005 TV189, do not posses uniform surface compositions. Further confirmation of both the model, and the classifications proposed by \citet{Fraser2012} will come from additional, and more accurate spectral variation measurements. In particular, it will be extremely interesting to test if the red class is one continuous class, as we have shown for the neutral class with TV189 and TD10, or further bifurcates into smaller subclasses. In addition, further measurements should be made to determine if objects do vary back and forth across the optical colour division between the red and neutral classes. If realized, this unexpected discovery would largely challenge the existence of two separate classes, and would require a broad rethink about our understanding of the compositions of small KBOs.
\acknowledgements

\bibliographystyle{apj}
\bibliography{astroelsart}

\begin{deluxetable}{lccccccc}
\tabletypesize{\scriptsize}
\tablecaption{Cycle 18 Spectral Photometry \label{tab:cycle18}}
\tablehead{\colhead{Designation\tablenotemark{a}} & \colhead{F606w} & \colhead{F775w} & \colhead{F110w} & \colhead{F098m} & \colhead{F127m} & \colhead{$H_{606}$} & \colhead{$\Delta m_{\textrm{606}}$}}
\setlength{\tabcolsep}{3pt}
\startdata
26308 (1998 SM165) & $21.89 \pm 0.01$ & $22.04 \pm 0.02$ & $22.20 \pm 0.01$ & $22.02 \pm 0.01$ & $22.32 \pm 0.01$ & $6.13 \pm 0.01$ & $-0.11 \pm 0.02$ \\
137295 (1999 RB216) & $22.99 \pm 0.03$ & $23.16 \pm 0.05$ & $23.83 \pm 0.03$ & $23.48 \pm 0.03$ & $24.02 \pm 0.04$ & $7.70 \pm 0.03$ & $0.26 \pm 0.04$ \\
29981 (1999 TD10) & $22.56 \pm 0.01$ & $22.75 \pm 0.02$ & $23.41 \pm 0.02$ & $23.06 \pm 0.02$ & $23.52 \pm 0.02$ & $9.34 \pm 0.01$ & $0.57 \pm 0.02$ \\
2000 OH67 & $23.86 \pm 0.04$ & $24.05 \pm 0.06$ & $24.55 \pm 0.07$ & $24.25 \pm 0.10$ & $24.77 \pm 0.08$ & $7.49 \pm 0.04$ & $-0.29 \pm 0.06$ \\
2000 WT169 & $22.92 \pm 0.02$ & $22.91 \pm 0.04$ & $23.48 \pm 0.03$ & $23.23 \pm 0.03$ & $23.71 \pm 0.03$ & $6.41 \pm 0.02$ & $0.09 \pm 0.03$ \\
2001 PK47 & $23.49 \pm 0.04$ & $23.57 \pm 0.06$ & $24.10 \pm 0.03$ & $23.74 \pm 0.06$ & $24.20 \pm 0.05$ & $7.78 \pm 0.04$ & $0.00 \pm 0.05$ \\
2001 QX297 & $23.13 \pm 0.02$ & $23.27 \pm 0.04$ & $23.86 \pm 0.04$ & $23.54 \pm 0.05$ & $24.00 \pm 0.04$ & $6.81 \pm 0.02$ & $-0.28 \pm 0.03$ \\
2001 QX322 & \nodata & \nodata & $23.88 \pm 0.04$ & $23.53 \pm 0.03$ & $24.01 \pm 0.04$ & \nodata & \nodata \\
2002 PV170 & $22.70 \pm 0.01$ & $22.76 \pm 0.02$ & $23.36 \pm 0.02$ & $23.00 \pm 0.03$ & $23.52 \pm 0.02$ & $6.49 \pm 0.01$ & $-0.11 \pm 0.02$ \\
120216 (2004 EW95) & $21.23 \pm 0.01$ & $21.57 \pm 0.01$ & $22.32 \pm 0.01$ & $21.97 \pm 0.01$ & $22.50 \pm 0.01$ & $6.80 \pm 0.01$ & $0.02 \pm 0.01$ \\
2004 QQ26 & $23.18 \pm 0.02$ & $23.38 \pm 0.04$ & $24.18 \pm 0.03$ & $23.90 \pm 0.06$ & $24.32 \pm 0.05$ & $10.1 \pm 0.02$ & $-0.06 \pm 0.03$ \\
2005 TV189 & $23.15 \pm 0.03$ & $23.40 \pm 0.04$ & $24.11 \pm 0.03$ & $23.85 \pm 0.04$ & $24.34 \pm 0.06$ & $8.12 \pm 0.03$ & $0.17 \pm 0.03$ \\
225088 (2007 OR10) & $21.66 \pm 0.01$ & $21.36 \pm 0.01$ & $21.699 \pm 0.007$ & $21.45 \pm 0.01$ & $21.826 \pm 0.008$ & $2.32 \pm 0.01$ & $-0.00 \pm 0.01$ \\
\enddata
\end{deluxetable}

\begin{deluxetable}{lcccccccc}
\tabletypesize{\scriptsize}
\tablecaption{Cycle 17 Spectral Photometry \label{tab:cycle17}}
\tablehead{\colhead{Designation\tablenotemark{a}} & \colhead{a (AU)} & \colhead{i ($^o$)} & \colhead{e} & \colhead{F606w} & \colhead{F814w} & \colhead{F139m} & \colhead{F153m} & \colhead{H$_{606}$}}
\setlength{\tabcolsep}{3pt}
\startdata
Nessus (1993 HA2) & 24.47 & 15.65 & 0.51 & $23.32 \pm 0.02$ & $23.18 \pm 0.02$ & $24.04 \pm 0.04$ & $24.39 \pm 0.04$ & $9.82 \pm 0.02$ \\
Hylonome (1995 DW2) & 24.99 & 4.14 & 0.24 & $23.41 \pm 0.03$ & $23.77 \pm 0.04$ & $24.60 \pm 0.07$ & $24.92 \pm 0.08$ & $10.12 \pm 0.04$ \\
26308 (1998 SM165) & 47.68 & 13.50 & 0.36 & $22.00 \pm 0.02$ & $22.06 \pm 0.02$ & $22.80 \pm 0.02$ & $22.99 \pm 0.02$ & $6.18 \pm 0.02$ \\
Cyllarus (1998 TF35) & 26.27 & 12.63 & 0.37 & $22.48 \pm 0.01$ & $22.42 \pm 0.01$ & $23.26 \pm 0.03$ & $23.56 \pm 0.03$ & $8.82 \pm 0.01$ \\
69986 (1998 WW24) & 39.63 & 13.92 & 0.22 & $23.86 \pm 0.05$ & $24.01 \pm 0.04$ & $24.88 \pm 0.13$ & $25.13 \pm 0.10$ & $8.68 \pm 0.04$ \\
1998 WY24 & 43.45 & 1.90 & 0.04 & $23.57 \pm 0.03$ & $23.70 \pm 0.04$ & $24.59 \pm 0.07$ & $24.86 \pm 0.08$ & $7.39 \pm 0.04$ \\
1999 CL119 & 47.15 & 23.24 & 0.01 & $23.06 \pm 0.02$ & $23.23 \pm 0.02$ & $24.39 \pm 0.06$ & $24.67 \pm 0.06$ & $6.40 \pm 0.02$ \\
40314 (1999 KR16) & 48.73 & 24.81 & 0.30 & $21.61 \pm 0.01$ & $21.52 \pm 0.01$ & $22.35 \pm 0.01$ & $22.69 \pm 0.01$ & $6.11 \pm 0.01$ \\
137295 (1999 RB216) & 47.58 & 12.69 & 0.29 & $22.72 \pm 0.01$ & $22.88 \pm 0.02$ & $23.92 \pm 0.03$ & $24.20 \pm 0.04$ & $7.49 \pm 0.02$ \\
1999 RJ215 & 59.23 & 19.73 & 0.41 & $23.06 \pm 0.02$ & $23.26 \pm 0.02$ & $24.14 \pm 0.06$ & $24.42 \pm 0.05$ & $7.71 \pm 0.02$ \\
86177 (1999 RY215) & 45.18 & 22.21 & 0.23 & $22.82 \pm 0.02$ & $23.24 \pm 0.02$ & $24.23 \pm 0.05$ & $24.41 \pm 0.06$ & $7.31 \pm 0.02$ \\
91554 (1999 RZ215) & 101.0 & 25.54 & 0.69 & $23.14 \pm 0.02$ & $23.38 \pm 0.03$ & $24.24 \pm 0.06$ & $24.54 \pm 0.08$ & $7.98 \pm 0.03$ \\
29981 (1999 TD10) & 97.79 & 5.95 & 0.87 & $21.99 \pm 0.01$ & $22.27 \pm 0.01$ & $23.20 \pm 0.03$ & $23.50 \pm 0.03$ & $9.04 \pm 0.01$ \\
Elatus (1999 UG5) & 11.81 & 5.24 & 0.38 & $22.45 \pm 0.01$ & $22.51 \pm 0.01$ & $23.68 \pm 0.04$ & $23.93 \pm 0.05$ & $10.93 \pm 0.01$ \\
121725 (1999 XX143) & 17.98 & 6.76 & 0.45 & $23.25 \pm 0.02$ & $23.39 \pm 0.03$ & $24.50 \pm 0.06$ & $24.79 \pm 0.06$ & $9.11 \pm 0.03$ \\
2000 AF255 & 48.89 & 30.87 & 0.24 & $23.37 \pm 0.02$ & $23.23 \pm 0.02$ & $23.96 \pm 0.04$ & $24.43 \pm 0.04$ & $6.30 \pm 0.02$ \\
2000 CE105 & 44.15 & 0.55 & 0.06 & $23.91 \pm 0.03$ & $23.92 \pm 0.04$ & $24.94 \pm 0.09$ & $25.06 \pm 0.13$ & $7.77 \pm 0.04$ \\
2000 CQ105 & 57.36 & 19.64 & 0.39 & $23.28 \pm 0.02$ & $23.68 \pm 0.03$ & $24.62 \pm 0.07$ & $24.97 \pm 0.10$ & $6.56 \pm 0.03$ \\
60608 (2000 EE173)\tablenotemark{b} & 49.61 & 5.94 & 0.54 & $21.83 \pm 0.1$ & $22.09 \pm 0.1$ & $23.32 \pm 0.02$ & $23.61 \pm 0.02$ & $8.2 \pm 0.1$ \\
60620 (2000 FD8) & 43.80 & 19.49 & 0.21 & $22.90 \pm 0.02$ & $22.91 \pm 0.02$ & $23.54 \pm 0.04$ & $23.94 \pm 0.03$ & $7.08 \pm 0.02$ \\
2000 FV53 & 39.18 & 17.33 & 0.16 & $23.30 \pm 0.02$ & $23.74 \pm 0.04$ & $24.64 \pm 0.07$ & $24.81 \pm 0.07$ & $8.06 \pm 0.04$ \\
130391 (2000 JG81) & 47.37 & 23.46 & 0.27 & $23.75 \pm 0.04$ & $24.00 \pm 0.04$ & $24.72 \pm 0.08$ & $25.02 \pm 0.09$ & $8.33 \pm 0.04$ \\
2000 OH67 & 44.02 & 5.62 & 0.01 & $24.15 \pm 0.04$ & $24.21 \pm 0.05$ & $25.05 \pm 0.15$ & $25.42 \pm 0.17$ & $7.79 \pm 0.05$ \\
87555 (2000 QB243) & 34.78 & 6.79 & 0.56 & $23.08 \pm 0.03$ & $23.30 \pm 0.03$ & $24.22 \pm 0.05$ & $24.44 \pm 0.06$ & $9.14 \pm 0.03$ \\
123509 (2000 WK183) & 44.61 & 1.96 & 0.04 & $23.13 \pm 0.03$ & $23.26 \pm 0.03$ & $24.25 \pm 0.04$ & $24.49 \pm 0.05$ & $6.84 \pm 0.03$ \\
2000 WT169 & 45.11 & 1.74 & 0.00 & $22.82 \pm 0.02$ & $22.94 \pm 0.02$ & $23.83 \pm 0.06$ & $24.14 \pm 0.06$ & $6.32 \pm 0.02$ \\
2000 YB2 & 38.90 & 3.81 & 0.02 & $23.06 \pm 0.02$ & $23.28 \pm 0.04$ & $24.07 \pm 0.06$ & $24.45 \pm 0.05$ & $7.21 \pm 0.04$ \\
2000 YH2 & 39.54 & 12.90 & 0.30 & $22.66 \pm 0.02$ & $23.01 \pm 0.02$ & $24.28 \pm 0.10$ & $24.40 \pm 0.05$ & $8.30 \pm 0.02$ \\
63252 (2001 BL41) & 9.725 & 12.49 & 0.29 & $22.99 \pm 0.03$ & $23.35 \pm 0.04$ & $24.05 \pm 0.04$ & $24.32 \pm 0.04$ & $12.06 \pm 0.04$ \\
150642 (2001 CZ31) & 45.48 & 10.19 & 0.12 & $22.21 \pm 0.01$ & $22.60 \pm 0.01$ & $23.60 \pm 0.03$ & $23.86 \pm 0.03$ & $6.13 \pm 0.01$ \\
131318 (2001 FL194) & 39.41 & 13.68 & 0.17 & $23.30 \pm 0.02$ & $23.65 \pm 0.03$ & $24.71 \pm 0.09$ & $24.81 \pm 0.10$ & $8.13 \pm 0.03$ \\
2001 FO185 & 46.53 & 10.63 & 0.12 & \nodata & \nodata & $23.99 \pm 0.06$ & $25.05 \pm 0.16$ & \nodata \\
82158 (2001 FP185) & 220.6 & 30.77 & 0.84 & \nodata & \nodata & $23.03 \pm 0.01$ & $23.33 \pm 0.02$ & \nodata \\
2001 FQ185 & 47.78 & 3.23 & 0.23 & $23.25 \pm 0.02$ & $23.16 \pm 0.02$ & $24.11 \pm 0.04$ & $24.34 \pm 0.04$ & $7.61 \pm 0.02$ \\
82155 (2001 FZ173) & 85.79 & 12.70 & 0.62 & \nodata & \nodata & $22.87 \pm 0.01$ & $23.17 \pm 0.01$ & \nodata \\
2001 HY65 & 43.28 & 17.12 & 0.12 & $22.64 \pm 0.01$ & $22.76 \pm 0.02$ & $23.76 \pm 0.04$ & $24.05 \pm 0.05$ & $6.69 \pm 0.02$ \\
2001 KA77 & 47.28 & 11.93 & 0.09 & $22.61 \pm 0.01$ & $22.55 \pm 0.01$ & $23.39 \pm 0.03$ & $23.67 \pm 0.03$ & $5.78 \pm 0.01$ \\
88269 (2001 KF77) & 25.91 & 4.35 & 0.23 & $23.84 \pm 0.04$ & $23.77 \pm 0.04$ & $24.51 \pm 0.12$ & $24.99 \pm 0.12$ & $10.71 \pm 0.04$ \\
88268 (2001 KK76) & 42.37 & 1.88 & 0.01 & $23.33 \pm 0.02$ & $23.32 \pm 0.03$ & $24.14 \pm 0.05$ & $24.38 \pm 0.06$ & $7.09 \pm 0.03$ \\
160147 (2001 KN76) & 43.68 & 2.64 & 0.08 & $23.33 \pm 0.02$ & $23.50 \pm 0.03$ & $24.56 \pm 0.10$ & $24.73 \pm 0.06$ & $7.33 \pm 0.03$ \\
2001 OQ108 & 45.52 & 2.32 & 0.01 & $23.82 \pm 0.03$ & $23.81 \pm 0.04$ & $24.67 \pm 0.10$ & $24.97 \pm 0.08$ & $7.26 \pm 0.04$ \\
2001 PK47 & 39.80 & 8.72 & 0.06 & $23.48 \pm 0.02$ & $23.78 \pm 0.05$ & $24.64 \pm 0.06$ & $24.76 \pm 0.06$ & $7.75 \pm 0.05$ \\
2001 QC298 & 46.19 & 30.62 & 0.12 & $22.63 \pm 0.02$ & $23.15 \pm 0.03$ & $24.15 \pm 0.05$ & $24.43 \pm 0.06$ & $6.59 \pm 0.03$ \\
139775 (2001 QG298) & 39.36 & 6.49 & 0.19 & $22.56 \pm 0.01$ & $22.62 \pm 0.01$ & $23.57 \pm 0.03$ & $23.85 \pm 0.04$ & $7.61 \pm 0.01$ \\
2001 QR297 & 44.25 & 5.15 & 0.02 & $23.47 \pm 0.03$ & $23.53 \pm 0.04$ & $24.39 \pm 0.07$ & $24.61 \pm 0.10$ & $7.07 \pm 0.04$ \\
2001 QR322 & 30.24 & 1.32 & 0.02 & $23.23 \pm 0.02$ & $23.56 \pm 0.03$ & $24.47 \pm 0.06$ & $24.76 \pm 0.06$ & $8.52 \pm 0.03$ \\
2001 QS322 & 43.96 & 0.24 & 0.03 & $23.64 \pm 0.03$ & $23.72 \pm 0.03$ & $24.72 \pm 0.10$ & $24.87 \pm 0.13$ & $7.40 \pm 0.03$ \\
Teharonhiawako & 43.89 & 2.58 & 0.02 & $22.856 \pm 0.02$ & $22.904 \pm 0.02$ & $24.020 \pm 0.07$ & $24.308 \pm 0.05$ & $6.47 \pm 0.02$ \\
(2001 QT297)\\
2001 QX297 & 44.01 & 0.91 & 0.02 & $23.42 \pm 0.02$ & $23.48 \pm 0.04$ & $24.39 \pm 0.08$ & $24.56 \pm 0.09$ & $7.06 \pm 0.04$ \\
2001 QX322 & 58.01 & 28.50 & 0.38 & $22.97 \pm 0.02$ & $23.17 \pm 0.02$ & $24.13 \pm 0.04$ & $24.46 \pm 0.05$ & $6.85 \pm 0.02$ \\
2001 RW143 & 43.06 & 2.96 & 0.03 & $23.91 \pm 0.04$ & $24.05 \pm 0.05$ & \nodata & \nodata & $7.73 \pm 0.05$ \\
2001 RZ143 & 44.12 & 2.12 & 0.06 & $23.03 \pm 0.02$ & $23.14 \pm 0.02$ & $24.07 \pm 0.06$ & $24.29 \pm 0.06$ & $6.93 \pm 0.02$ \\
119315 (2001 SQ73) & 17.50 & 17.43 & 0.17 & $22.04 \pm 0.01$ & $22.34 \pm 0.01$ & $23.31 \pm 0.03$ & $23.54 \pm 0.03$ & $9.65 \pm 0.01$ \\
2001 UP18 & 47.79 & 1.17 & 0.07 & \nodata & \nodata & $24.53 \pm 0.06$ & $24.82 \pm 0.07$ & \nodata \\
148975 (2001 XA255) & 28.92 & 12.62 & 0.67 & $21.48 \pm 0.01$ & $21.78 \pm 0.01$ & $22.77 \pm 0.02$ & $22.97 \pm 0.02$ & $11.72 \pm 0.01$ \\
126155 (2001 YJ140) & 39.74 & 5.96 & 0.29 & $22.22 \pm 0.01$ & $22.49 \pm 0.01$ & $23.35 \pm 0.02$ & $23.57 \pm 0.02$ & $7.78 \pm 0.01$ \\
2002 CU154 & 44.02 & 3.34 & 0.06 & $23.66 \pm 0.03$ & $23.72 \pm 0.04$ & $24.67 \pm 0.06$ & $25.07 \pm 0.08$ & $7.56 \pm 0.04$ \\
2002 FW36 & 43.09 & 2.35 & 0.02 & \nodata & \nodata & $24.90 \pm 0.11$ & $25.11 \pm 0.13$ & \nodata \\
Crantor (2002 GO9) & 19.35 & 12.79 & 0.27 & $21.09 \pm 0.01$ & $21.16 \pm 0.01$ & $21.83 \pm 0.01$ & $22.13 \pm 0.01$ & $9.18 \pm 0.01$ \\
2002 GY32 & 39.44 & 1.80 & 0.09 & $23.32 \pm 0.02$ & $23.40 \pm 0.03$ & $24.36 \pm 0.05$ & $24.60 \pm 0.06$ & $7.78 \pm 0.03$ \\
2002 PD155 & 43.14 & 5.78 & 0.00 & $23.99 \pm 0.04$ & $24.14 \pm 0.07$ & $24.80 \pm 0.14$ & $25.42 \pm 0.16$ & $7.70 \pm 0.07$ \\
2002 PV170 & 42.55 & 1.27 & 0.01 & $22.81 \pm 0.02$ & $22.88 \pm 0.02$ & $23.85 \pm 0.04$ & $24.09 \pm 0.04$ & $6.54 \pm 0.02$ \\
2002 QX47 & 25.38 & 7.26 & 0.37 & \nodata & \nodata & $23.42 \pm 0.02$ & $23.67 \pm 0.02$ & \nodata \\
119976 (2002 VR130) & 24.07 & 3.52 & 0.38 & $23.47 \pm 0.03$ & $23.88 \pm 0.04$ & $24.65 \pm 0.07$ & $25.04 \pm 0.08$ & $11.72 \pm 0.04$ \\
119979 (2002 WC19) & 48.19 & 9.16 & 0.26 & $21.151 \pm 0.009$ & $21.24 \pm 0.01$ & $22.10 \pm 0.01$ & $22.38 \pm 0.01$ & $4.88 \pm 0.009$ \\
127546 (2002 XU93) & 67.50 & 77.89 & 0.68 & $21.71 \pm 0.01$ & $22.08 \pm 0.01$ & $23.07 \pm 0.01$ & $23.29 \pm 0.01$ & $8.49 \pm 0.01$ \\
2003 FE128 & 47.95 & 3.38 & 0.25 & $22.31 \pm 0.01$ & $22.39 \pm 0.02$ & $23.38 \pm 0.02$ & $23.70 \pm 0.02$ & $6.81 \pm 0.02$ \\
2003 FF128 & 39.47 & 1.91 & 0.21 & $22.34 \pm 0.01$ & $22.38 \pm 0.01$ & $23.38 \pm 0.03$ & $23.63 \pm 0.02$ & $7.31 \pm 0.01$ \\
Ceto (2003 FX128) & 101.9 & 22.27 & 0.82 & $21.85 \pm 0.01$ & $22.07 \pm 0.01$ & $22.95 \pm 0.01$ & $23.25 \pm 0.01$ & $6.89 \pm 0.01$ \\
385437 (2003 GH55) & 44.16 & 1.10 & 0.08 & $22.63 \pm 0.01$ & $22.69 \pm 0.02$ & $23.49 \pm 0.03$ & $23.80 \pm 0.04$ & $6.53 \pm 0.02$ \\
2003 HG57 & 43.67 & 2.09 & 0.03 & $23.04 \pm 0.03$ & $23.31 \pm 0.05$ & $23.84 \pm 0.11$ & $24.04 \pm 0.12$ & $5.79 \pm 0.05$ \\
2003 LD9 & 47.12 & 6.98 & 0.17 & $23.38 \pm 0.02$ & $23.36 \pm 0.03$ & $24.41 \pm 0.05$ & $24.61 \pm 0.06$ & $7.23 \pm 0.03$ \\
2003 QA92 & 38.12 & 3.43 & 0.05 & $22.60 \pm 0.01$ & $22.69 \pm 0.02$ & $23.56 \pm 0.02$ & $23.83 \pm 0.03$ & $6.90 \pm 0.02$ \\
385447 (2003 QF113) & 43.67 & 4.46 & 0.02 & $23.60 \pm 0.03$ & $23.65 \pm 0.03$ & $24.66 \pm 0.10$ & $24.88 \pm 0.08$ & $7.34 \pm 0.03$ \\
2003 QX91 & 43.60 & 27.71 & 0.24 & $24.31 \pm 0.05$ & $24.56 \pm 0.10$ & $25.17 \pm 0.19$ & $25.51 \pm 0.20$ & $9.1 \pm 0.10$ \\
149560 (2003 QZ91) & 41.37 & 34.86 & 0.47 & $23.20 \pm 0.03$ & $23.43 \pm 0.03$ & $24.73 \pm 0.07$ & $24.90 \pm 0.08$ & $8.87 \pm 0.03$ \\
2003 SQ317 & 42.62 & 28.60 & 0.07 & $22.73 \pm 0.01$ & $23.29 \pm 0.04$ & $24.56 \pm 0.09$ & $26.12 \pm 0.26$ & $6.84 \pm 0.04$ \\
2003 UZ117 & 44.42 & 27.42 & 0.13 & $21.42 \pm 0.01$ & $21.94 \pm 0.01$ & $23.41 \pm 0.03$ & $24.63 \pm 0.08$ & $5.47 \pm 0.01$ \\
Sedna (2003 VB12) & 532.7 & 11.92 & 0.85 & \nodata & \nodata & $22.000 \pm 0.009$ & $22.31 \pm 0.01$ & \nodata \\
2003 WU172 & 39.48 & 4.14 & 0.26 & $21.35 \pm 0.01$ & $21.52 \pm 0.01$ & $22.43 \pm 0.01$ & $22.69 \pm 0.01$ & $6.67 \pm 0.01$ \\
2004 EH96 & 39.55 & 3.13 & 0.28 & $23.22 \pm 0.02$ & $23.19 \pm 0.02$ & $24.02 \pm 0.04$ & $24.41 \pm 0.05$ & $8.59 \pm 0.02$ \\
120216 (2004 EW95) & 39.44 & 29.27 & 0.31 & $21.20 \pm 0.01$ & $21.60 \pm 0.01$ & $22.68 \pm 0.01$ & $22.96 \pm 0.01$ & $6.80 \pm 0.01$ \\
90568 (2004 GV9) & 42.00 & 22.00 & 0.07 & $20.272 \pm 0.008$ & $20.434 \pm 0.007$ & $21.418 \pm 0.007$ & $21.718 \pm 0.009$ & $4.374 \pm 0.008$ \\
2004 PA112 & 38.88 & 32.97 & 0.10 & $23.23 \pm 0.02$ & $23.58 \pm 0.03$ & $24.38 \pm 0.08$ & $24.69 \pm 0.08$ & $7.79 \pm 0.03$ \\
2004 QQ26 & 22.91 & 21.45 & 0.14 & $23.24 \pm 0.02$ & $23.60 \pm 0.03$ & $24.55 \pm 0.07$ & $24.83 \pm 0.07$ & $10.16 \pm 0.03$ \\
2004 TV357 & 48.00 & 9.74 & 0.28 & $22.54 \pm 0.01$ & $23.04 \pm 0.02$ & $24.03 \pm 0.04$ & $24.32 \pm 0.04$ & $7.02 \pm 0.02$ \\
2004 VN112 & 334.0 & 25.52 & 0.85 & \nodata & \nodata & $24.85 \pm 0.17$ & $24.94 \pm 0.12$ & \nodata \\
2004 XR190 & 57.74 & 46.56 & 0.10 & $22.18 \pm 0.01$ & $22.46 \pm 0.01$ & $23.46 \pm 0.02$ & $23.73 \pm 0.02$ & $4.55 \pm 0.01$ \\
2005 CA79 & 48.18 & 11.66 & 0.22 & $21.298 \pm 0.009$ & $21.49 \pm 0.01$ & $22.47 \pm 0.02$ & $22.70 \pm 0.01$ & $5.606 \pm 0.009$ \\
2005 EZ296 & 39.55 & 1.77 & 0.15 & $22.93 \pm 0.03$ & $22.92 \pm 0.02$ & $23.88 \pm 0.05$ & $24.23 \pm 0.06$ & $7.44 \pm 0.02$ \\
2005 GB187 & 39.46 & 14.68 & 0.23 & $22.30 \pm 0.01$ & $22.65 \pm 0.02$ & \nodata & \nodata & $7.46 \pm 0.02$ \\
2005 GE187 & 39.27 & 18.25 & 0.32 & $22.60 \pm 0.02$ & $22.58 \pm 0.01$ & $23.38 \pm 0.02$ & $23.63 \pm 0.02$ & $7.85 \pm 0.01$ \\
2005 GF187 & 39.39 & 3.90 & 0.25 & $23.09 \pm 0.02$ & $23.49 \pm 0.03$ & $24.67 \pm 0.09$ & $24.79 \pm 0.07$ & $8.30 \pm 0.03$ \\
2005 PU21 & 175.5 & 6.17 & 0.83 & \nodata & \nodata & $24.13 \pm 0.04$ & $24.43 \pm 0.08$ & \nodata \\
160427 (2005 RL43) & 24.55 & 12.26 & 0.04 & $22.05 \pm 0.01$ & $22.03 \pm 0.01$ & $22.85 \pm 0.02$ & $23.13 \pm 0.01$ & $8.30 \pm 0.01$ \\
145452 (2005 RN43) & 41.39 & 19.28 & 0.02 & $20.166 \pm 0.006$ & $20.350 \pm 0.007$ & $21.377 \pm 0.007$ & $21.651 \pm 0.008$ & $4.048 \pm 0.006$ \\
2005 RO43 & 29.03 & 35.46 & 0.52 & $21.67 \pm 0.01$ & $21.97 \pm 0.01$ & $22.79 \pm 0.01$ & $23.07 \pm 0.01$ & $7.54 \pm 0.01$ \\
308379 (2005 RS43) & 47.78 & 10.00 & 0.19 & $21.72 \pm 0.01$ & $22.04 \pm 0.01$ & $23.11 \pm 0.02$ & $23.39 \pm 0.02$ & $5.46 \pm 0.01$ \\
145474 (2005 SA278) & 93.27 & 16.28 & 0.64 & $22.78 \pm 0.02$ & $23.09 \pm 0.02$ & $24.23 \pm 0.07$ & $24.46 \pm 0.05$ & $6.79 \pm 0.02$ \\
145480 (2005 TB190) & 75.40 & 26.50 & 0.38 & $21.328 \pm 0.009$ & $21.51 \pm 0.01$ & $22.58 \pm 0.01$ & $22.89 \pm 0.01$ & $4.708 \pm 0.009$ \\
2005 TV189 & 39.63 & 34.35 & 0.19 & $22.97 \pm 0.02$ & $23.25 \pm 0.02$ & $24.07 \pm 0.04$ & $24.32 \pm 0.04$ & $7.98 \pm 0.02$ \\
2005 VJ119 & 35.28 & 6.95 & 0.68 & $22.15 \pm 0.01$ & $22.53 \pm 0.03$ & $23.27 \pm 0.03$ & $23.63 \pm 0.04$ & $11.48 \pm 0.03$ \\
2006 HW122 & 45.44 & 1.53 & 0.06 & $23.99 \pm 0.04$ & $24.15 \pm 0.08$ & $24.91 \pm 0.19$ & $25.21 \pm 0.16$ & $7.56 \pm 0.08$ \\
2006 QH181 & 67.07 & 19.08 & 0.42 & $23.78 \pm 0.04$ & $23.82 \pm 0.04$ & $24.46 \pm 0.06$ & $24.65 \pm 0.08$ & $4.73 \pm 0.04$ \\
2006 QP180 & 38.01 & 4.95 & 0.65 & $22.72 \pm 0.02$ & $22.62 \pm 0.01$ & $23.30 \pm 0.03$ & $23.65 \pm 0.03$ & $10.15 \pm 0.01$ \\
308933 (2006 SQ372) & 755.4 & 19.47 & 0.96 & $21.95 \pm 0.01$ & $22.00 \pm 0.01$ & $22.93 \pm 0.02$ & $23.22 \pm 0.02$ & $8.13 \pm 0.01$ \\
187661 (2007 JG43) & 23.90 & 33.14 & 0.40 & $21.109 \pm 0.009$ & $21.43 \pm 0.01$ & $22.31 \pm 0.01$ & $22.38 \pm 0.02$ & $9.459 \pm 0.009$ \\
2007 JK43 & 46.14 & 44.91 & 0.48 & $21.254 \pm 0.009$ & $21.46 \pm 0.01$ & $22.32 \pm 0.01$ & $22.61 \pm 0.01$ & $7.472 \pm 0.009$ \\
225088 (2007 OR10) & 66.84 & 30.92 & 0.50 & $21.66 \pm 0.01$ & $21.41 \pm 0.01$ & $22.05 \pm 0.01$ & $22.47 \pm 0.01$ & $2.33 \pm 0.01$ \\
2007 RH283 & 15.93 & 21.33 & 0.33 & $21.407 \pm 0.009$ & $21.73 \pm 0.01$ & $22.72 \pm 0.01$ & $23.01 \pm 0.01$ & $8.846 \pm 0.009$ \\
2007 TA418 & 72.88 & 21.95 & 0.50 & $23.29 \pm 0.02$ & $23.50 \pm 0.03$ & $24.37 \pm 0.08$ & $24.66 \pm 0.08$ & $7.66 \pm 0.03$ \\
\enddata
\tablenotetext{a}{MPC designations or names and asteroid numbers (where available). H$_{606}$ is the absolute magnitude of the object determined from its observed magnitude in the F606w filter and its distance at time of observation.}
\tablenotetext{b}{F606w and F814w photometry determined from observations presented by \citet{Benecchi2011} (see Fraser~and~Brown~(2012)).}
\end{deluxetable}

\end{document}